# The Archaeo-Astronomy Project - Supporting the Outdoor Classroom

**D Brown[1], R Francis[2], and A Alder[3]**
[1] School of Science and Technology
Nottingham Trent University
Erasmus Darwin Building
Clifton campus
Nottingham
NG11 8NS
[2] School of Education
Nottingham Trent University
Ada Byron King Building
Nottingham Trent University
Clifton
Nottingham
NG11 8NS
[3] School of Animal, Rural, and Environmental Sciences
Nottingham Trent University
Brackenhurst campus
Southwell
Nottinghamshire
NG25 0QF

E-mail: daniel.brown02@ntu.ac.uk

**Abstract.** Field trips and the outdoor classroom are a vital part of many courses at school and FE/HE level. They apply previously obtained skills and knowledge in the field and further deeper learning. However, it would be desirable to teach the material in situ rather than providing a field trip at the end of a course; thereby avoiding referring to the field trip in a few months time where everything will become much clearer.

Furthermore, there is clear evidence provided by the National Peak Park Authority on how neglected e.g. the Peak District National Park sites are by Primary and Secondary schools. They quote inaccessibility and other common barriers associated to the outdoor classroom.

The archaeo-astronomy project envisages the development of an Elearning environment allowing FE/HE students and pupils (Key Stage 2-4) to experience and explore ancient landscapes e.g. in the Peak District National Park. This project would support the learning experience of students on science and environmental modules. Furthermore, it would allow schools to overcome the initial problems of outdoor classroom.

It is a research-based project located at Nottingham Trent University that spans a large age range from secondary school pupils up to undergraduate students. Additionally, it is a collaboration between different departments and highly interdisciplinary including aspects of Astronomy, Physics, Ecology, and Archaeology. A high priority is placed upon the feedback between participants and teachers, especially bridging the gap from HE to secondary schools and colleges. Thereby, students represent role models and support the learning and teaching experience at school level.

## 1. Introduction

In recent years the importance of including field trips into the school curriculum has become more widely recognized (Braud and Reiss, 2004). Especially, after the publication of the *Manifesto for out of classroom learning* (Department for education and skills, 2006) and its repeated mention in the Rose report (Rose, 2009), there has been a strong motivation from the government to include outdoor learning. Educational research has shown that using the outdoor as a learning and teaching environment furthers deeper learning, demonstrates the practicality of taught skills, and establishes many other transferable skills (see review by Rickinson *et al.* 2004). An innovative example is the so called forest schools (see Murray and O'Brien 2005). Learning and teaching activities based on the outdoor classroom have already been included in many FE/HE courses related to landscape subjects and into other subjects such as physics. Unterman (2001) illustrates the use of an amusement park as an outdoor physics classroom. Their potential has been analysed on school level by Anderson, Nielson and Nashon (2009) and extended to undergraduate teaching by Bagge and Pendrill (2002). Such work has only been carried out for astronomy on school level (see e.g. Asher et al. 2006 or Brown and Neale 2010).

The ideal situation in using the outdoor class room to its full potential would be to carry out all of the learning and teaching activities outside the traditional class room setting. However, this is rarely possible and a more realistic solution has to be found. Both schools and FE/HE institutions introduce the outdoor classroom to consolidate previously taught material and include a field trip at the end of a module. Therefore, teachers and lecturers frequently have to refer the students to the upcoming trip: where everything will become much clearer. This is clearly not ideal, since the opportunity for deeper learning is reduced by separating the acquiring of knowledge from the applications of skills.

Furthermore, schools struggle implementing the outdoor class room in the first place. They see Health and Safety or unclear links to the curriculum as barriers that are very difficult to overcome. An example for this can be found in the use of ancient monuments and their surrounding landscapes in the Peak District by local primary and secondary schools. *"… Stanton-in-Peak is probably the only local primary school to make practical use of the moor."* (McGuire and Smith, 2007). This situation is worsened, since initiatives that could support the use of such Peak Park sites in schools (e.g. Losehill Hall, Peak District National Park) do not encounter a demand given the schools reluctance to engage in such activities. This vicious circle can only be broken by creating a demand for such activities in schools and, therefore, schools seeing the potential in such activities.

In a first step we have outlined the potential for secondary schools of an archaeo-astronomy summer school utilising two well known ancient monuments in the peak park; and how easy it can be organized and integrated into the curriculum (Brown, Neale, and Francis 2010). The following will elaborate on a general project that aims to provide much more structured support promoting the outdoor classroom use of ancient monuments in the Peak District National Park. First we will summarize the intended outcomes of the project. This will be followed by a description of our innovative methods of carrying out the project, and an analysis of the impact we have had so far, given that our work is still in progress. Finally, we will discuss our findings and draw together our findings.

**2. Project Outcomes**

The outcomes of the archaeo-astronomy project are twofold and will initially target a specific area in the Peak District National Park known as Gardom's Edge.

First we intend to support the FE/HE teaching by creating a virtual tour of the history rich landscape of Gardom's Edge. Much teaching is based on this area, but visits or field trips tend to take place in the Easter period, because of varying factors such as weather and vernal plants.

The richness of this area covers the following intended teaching and learning outcomes:

- Landuse: The enclosure periods, millstone industry (with obvious links to geology), and upland sheep farming.
- Historical Landscapes: Ranging from Neolithic burials, rock art, and enclosures to medieval quarries, hermits, and salt ways.
- Moor land habitat management and Landscape Character Assessment.
- View points in the area provide access to a wealth of learning, such as reversed S shaped fields, planned countryside, and the use of visual observation as a management tool.

It would be beneficial to have an interactive landscape for students to explore alongside lectures and to prepare for field trips. The student understanding across a wide range of subject areas would be supported. One such example could be a virtual walk through a hollow way or through an enclosure area that would complement map work. Another could be how rock art and other Neolithic remains relate to the landscape. Other examples are class room based Landscape Character Assessment and comparisons.

A common comment in lectures is "*…we will show you examples of this on the field trip…*" whilst still in the classroom. Given the proposed environment, the students could enter the virtual landscape at will and then build on these explorations during the field trip.

Additionally, the resources developed will be used to allow schools to carry out virtual field trips with their class. They will be able to use the developed elearning environment to support material covered during the lessons:

- A virtual tour along a rock face to explore aspects of the rock cycle as part of the geological activity taught at KS 3 (Science 3.4 a). This also includes how mankind has changed the rocks according to the different uses covered in physical and human processes, environmental interaction at KS3 (Geography 1.5 a).
- Aspects of citizenship at KS3 (Citizenship 1.2b) will be targeted by introducing identities and diversities through interviews with people that used to use the landscape in the past millennia.
- The example of a standing stone close to a Neolithic enclosure will be used in an interactive panoramic viewing tool as well as a JAVA applet to support learning of Light and Shadow covered in KS2 (Sc4 3b), Seasons covered in KS2 (Sc4 3b), and consolidating the apparent motion of the Sun KS3 (Science 3.4b).

Teachers will also be given resources for different science projects that can be carried out either on site or in their school grounds with small groups, e.g. Science Clubs or Gifted and Talented (G and T) pupils, further supporting the inclusion of the outdoor classroom into the curriculum.

## 3. Research and Project Development Methods

Given limited resources we decided to subdivide the project into several mini-projects. These could then be carried out over a longer period of time with a smaller amount of staff. Initially this appears to reduce the effectiveness of developing the project. However, this approach allowed us to integrate students from schools and undergraduates into our work. As a result, we could trial some of our activities and collect feedback on the suitability of the resources, some of which were developed by the students themselves. The most important aspect of this approach was the support of students in the transitional period from school to FE/HE. We achieved this by introducing project based learning to work experience students and allowing them to experience a more realistic learning environment comparable to universities. Students also worked together with undergraduates, PhD students and lecturers providing valuable results and thereby increasing their self confidence. Their work also led on to publications and resources used by either their peers or other institutions. A brief schematic overview of our approach is shown in figure 1. The initial separation of schools and FE/HE is removed by linking key participants from all areas into joint projects or into resource development, e.g. for master classes. The following four examples will outline briefly how they support transition and increase self confidence.

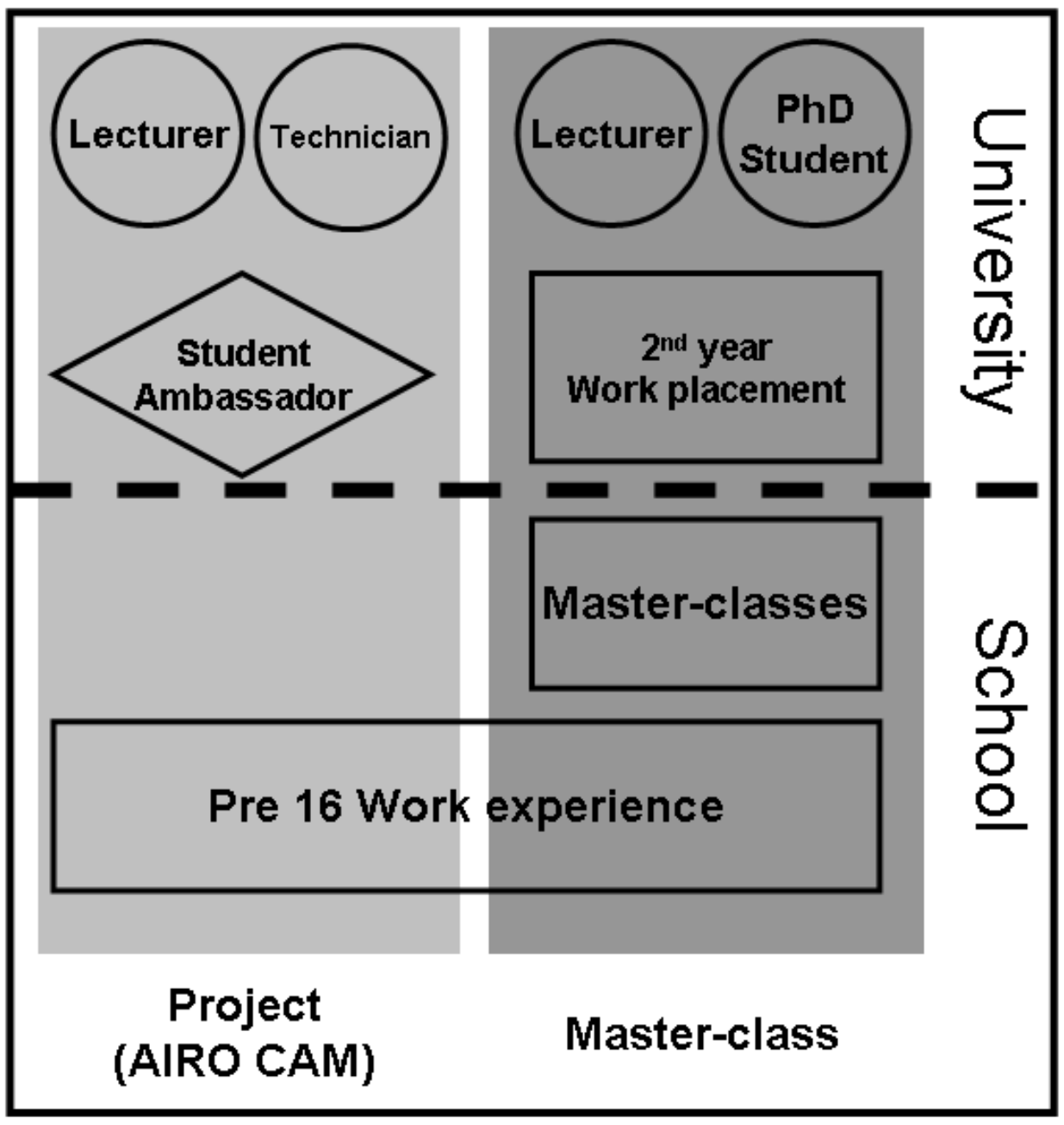

**Figure 1** – Impact of subdividing overall project development into mini-projects, e.g. AIRO CAM, on supporting transition from schools to universities. Members of different educational institutes become inter-dependant within a mini-project easing transition.

*3.1 Panoramas and Stellarium*

Observing the night sky in a school environment has always involved flexibility of teaching staff, students, and their parents. However, planetarium programs such as Stellarium[1] are now freely available and very easy to use. This program can visualise key concepts of astronomy taught at school. It allows generating and developing excellent visual resources that are essential for our overall goal. A group of G and T year 8 students created a continuing personal development (CPD) resource set (WIKI page[2], video, and brochure) for secondary school teachers on the example of using the North star for orienteering.

Furthermore, based on the work of the G and T students a work placement FdSc 2nd year Bioscience student at Nottingham Trent University acquired panoramic imaging from megalithic sites in the Peak District National Park and included them into Stellarium as shown in figure 2 to compare the horizon profile with specific lunar and solar alignments (more details on resources WIKI page[3]). This work has inspired many of our work experience students and led to work on three dimensional panoramas in Stellarium. Most importantly it will be used by the National Peak Park Authority Dark Skies project to raise the public awareness towards light pollution in the context of the nocturnal environment as seen in the example of Nine Stone Close in figure 3.

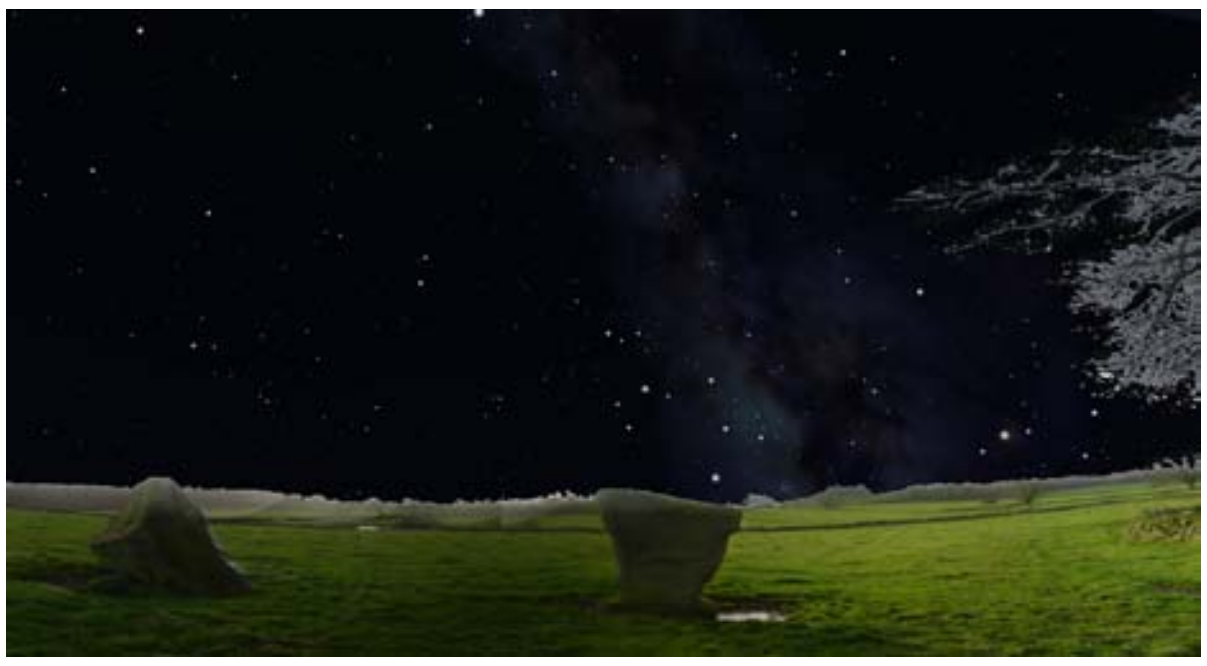

**Figure 2** – The view of the night sky created by Stellarium and a panoramic landscape of Nine Stone Close. This site can be explored in its appropriate nocturnal environment.

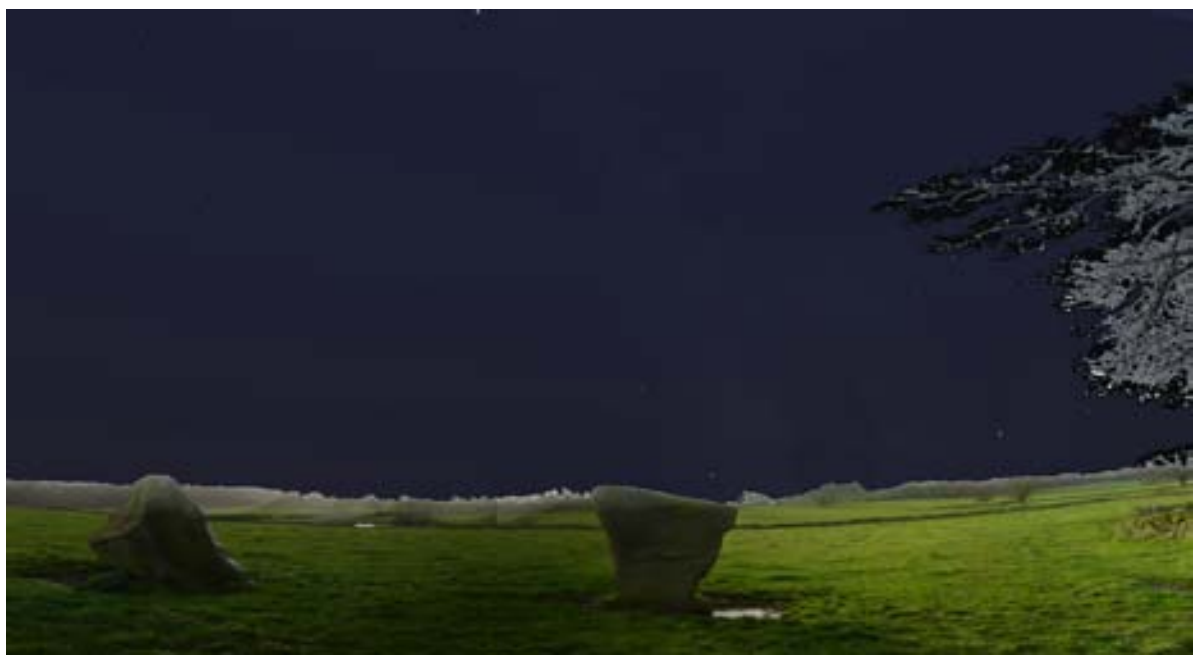

**Figure 3** – As in Figure 2 but including light pollution. This demonstrates how many stars we lose as well as destroying a beautiful nocturnal landscape.

[1] Stellarium is a free planetarium software available at http://www.stellarium.org/ .
[2] A detailed tutorial and materials can be found at http://medenschool.pbworks.com/Stellarium.
[3] Stellarium resources are available at http://medenschool.pbworks.com/resources.

*3.2 Arial Imaging and Site Documentation*

Our proposed Elearning environment has to be able to visualise an outdoor classroom environment. To include realistic images especially, we established several projects that either collected aerial images of ancient monuments or documented their surroundings.

We developed an aerial imaging unit (AIRO CAM[4]) that was built and field tested by work experience students. They also gathered data and created aerial views of an ancient megalithic monument (Arbor Low) presented in figure 4 comparable in precisions to archaeological site plans by Barnatt (1990).

Further documentation of ancient sites was carried out by a placement student and a member of the team using standard photography and video recordings. The accessibility of the site and its educational added value were investigated.

The aspects of accessibility and added educational opportunities were covered during the documentation of an ancient megalithic monument (Minning Low). These include evidence of modern ritual forming a discussion for Personal, Social, Health and Economic or Religious Education classes.

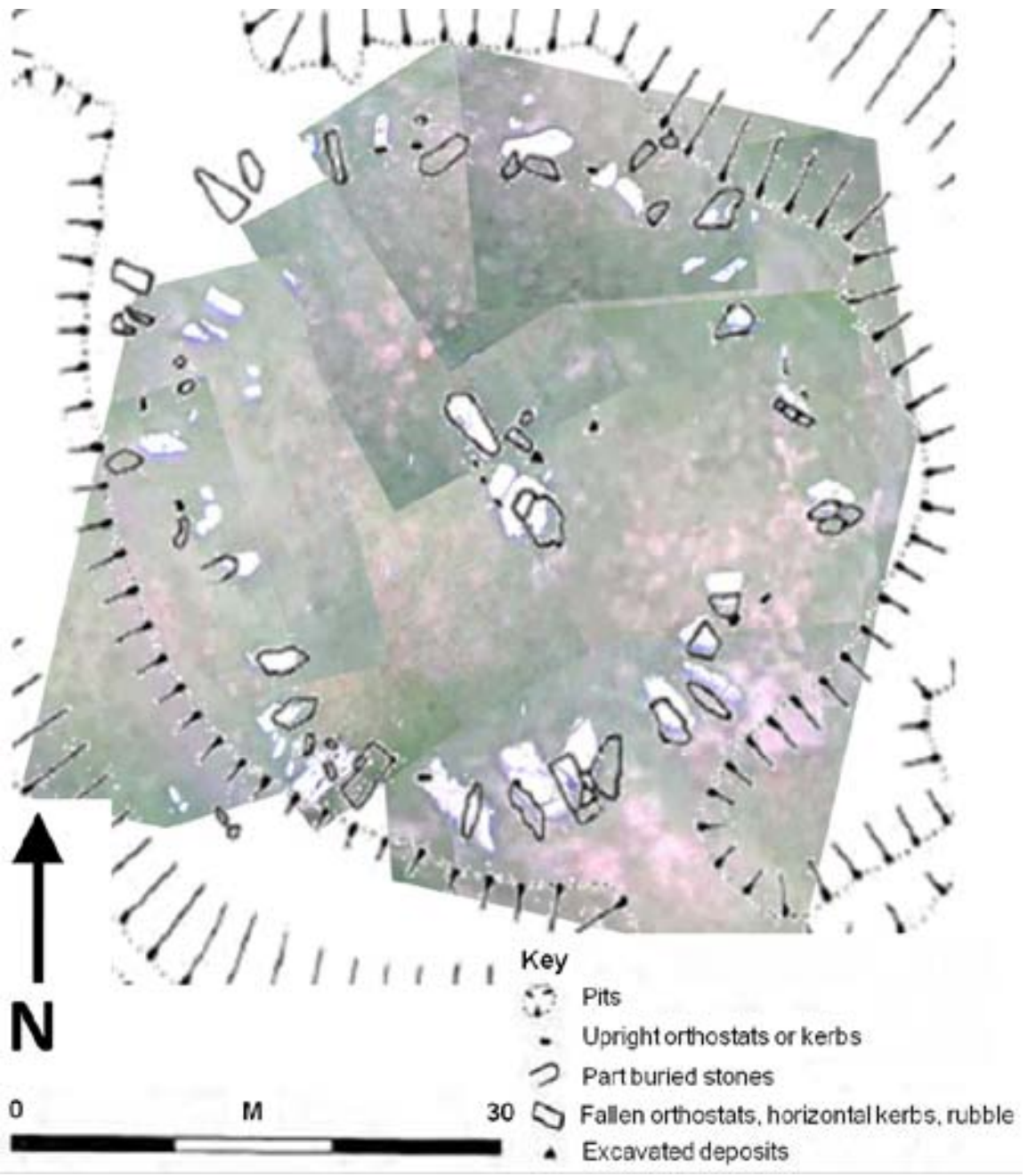

**Figure 4** – The mosaic of an aerial view of the Arbor Low stone circle taken with the AIRO CAM unit. The overlay is an adaptation of a site plan by Barnatt (1990). The precision of the AIRO CAM data is comparable in quality and information.

*3.3 Summer School*

The summer school allowed the team to develop and trial several new resources, workshops, and activities that would link into the early years of the secondary curriculum. The team were then able to reflect upon their experience and include improvements. Using a cross curriculum strategy, multiple subjects were targeted during the same activity. An activity based on archaeological excavations would use both elements of history and geography to explain relative dating as stratigraphy. A more detailed description of the summer school can be found in Brown, Neale, and Francis (2010).
The feedback from teachers and students was positive and e.g. led to improvements with the “virtual voyage through time” activity. Here students imagine the landscape of Stone Age Britain while a member of the team narrates the story of how the first hunter gatherers arrived and began to settle. Future improvements will add a full Visual-Auditory-Reading-Kinaesthetic (VARK) experience including sound effects, physical sensations, and smell. A similar holistic experience has also been successfully offered to visitors at the Jorvik Viking Centre in York (Aggleton *et al.* 1999).

[4] Resources and results on AIROCAM can be found at http://archaeopoject.pbworks.com/AIRO-CAM.

*3.4 Master Classes*

We have developed a set of master classes we offer to secondary schools. One of them includes an archaeo-astronomy tour through the landscape surrounding Gardom’s Edge. During this guided tour the students are involved in very short workshops on different aspects of the landscape, including a rare isolated standing stone erected during the Neolithic period. We have used this location in the past to elaborate on the practicality of fore- and back-sights for alignments in stone circles. Together with a work experience student, a university technician and lecturer during June 2010 we surveyed the stone more precisely and discovered a possible use as a seasonal marker. Note the dip of the slope in figure 5 being close to the Sun’s altitude during the Summer solstice. The initial survey by the student was followed up by a PhD student working on rock erosion.

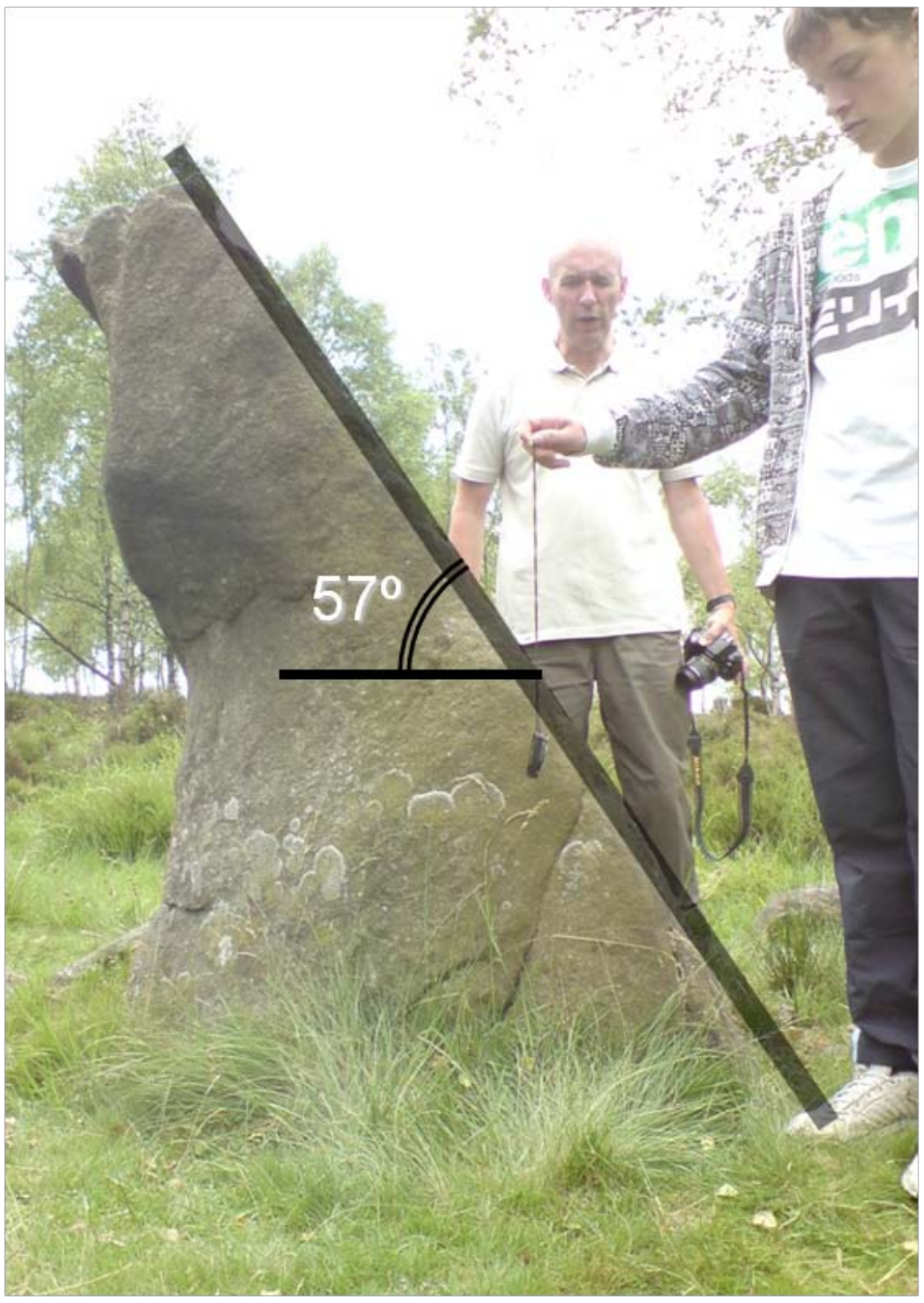

**Figure 5** – Initial survey of standing stone at Gardom’s Edge carried out by pre-16 work experience student. The slope of the stone indicates a possible seasonal sundial, since it is close to the Sun’s altitude during the Summer solstice (~60°).

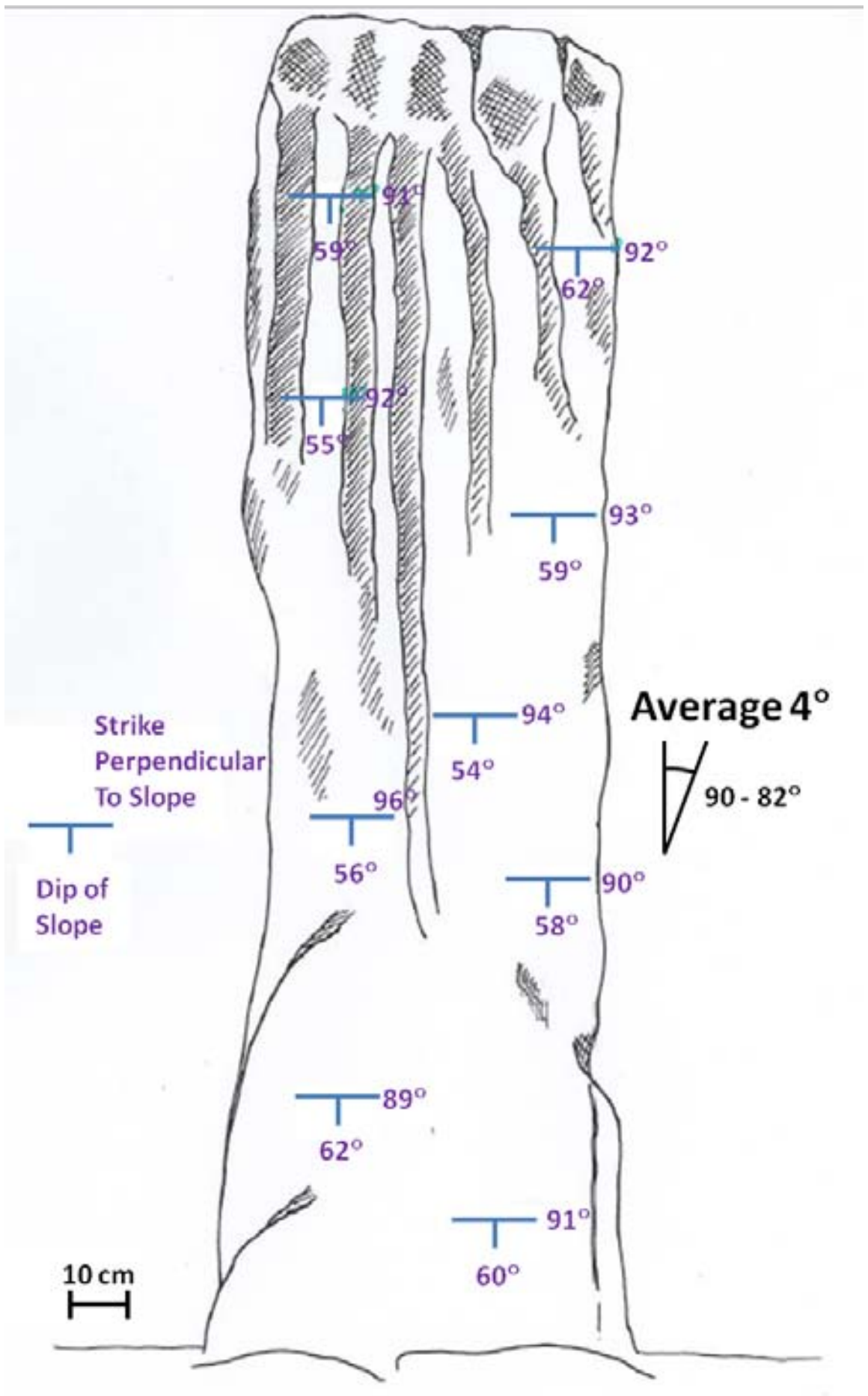

**Figure 6** –Follow-up work carried out by a PhD student at the standing stone to survey the weathering. Measurements indicate the strike and pitch of the North facing slope. (Sketch reproduced with permission of E. Bemand.)

She carried out a more detailed study, part of which is a detailed map of the dip of the slope in figure 6 that now strengthens our seasonal marker theory (see Brown, Bemand and Alder 2010). As a result, we now have embraced this interpretation and added a stronger link to seasons and the Sun's apparent motion during the day and year.

**Analysis**

Although a complete Elearning environment supporting the outdoor classroom has not been achieved yet, a basic WIKI environment has been set up containing resources and imaging results (see archaeopoject.pbworks.com). We will analyse our preliminary results regarding the support of transition from schools to FE/HE institutes below.

Especially the popularity of our activities we offer to schools in form of master classes and work experience placements demonstrates the success our project has had so far. All our four placements offered in the period from 2009/2010 were filled, and within the first two months of period 2010/2011 the next four placements have been filled. This has led us to offer an additional four places for that period that have now already been overbooked twice over. Furthermore, we have had approximately 60 students taking part in archaeo-astronomy master classes over the past year. And only recently our project led to the establishment of two CPD activities on the use of ancient sites as outdoor classroom for 80 local teachers in the East Midlands as well as 13 international Korean teachers.

Apart from the popularity, the feedback we have received has also been very good. Work experience students have commented that the placement has been "*Fantastic*", "*It has been brilliant.*" or they "*...would thoroughly recommend it to anyone...*". They frequently name the field trip or the AIRO CAM project as their biggest achievement. Even if a student was not involved in field testing but developed lab based experiments with AIRO CAM he remarked: "*I found the week really interesting, especially (…) doing the experiments* (with AIRO CAM) *in the Lab*.". These anecdotal findings can be supported by the results from the student evaluation form. This form contained a section in which students had to rate their placement on a Likert scale, regarding e.g. the tasks they were given, the support they received, and the experience they had of the workplace and university. The results are shown in figure 7, whereby the experience is an average between workplace and university. Although this is a very small sample of only six students, it still shows that they all enjoyed the tasks and always felt they had adequate support. Bearing in mind this support was provided through resources developed by their peers or undergraduates and personal support by lecturers and technicians. Most importantly, they all thought this was an ideal introduction into the university learning environment.

Finally, the skills and knowledge obtained by participant during any stage of the project was longer retained and they showed clear signs of a deeper understanding. This can be seen in the high standard and usefulness of resources in the shape of web based WIKI pages, video tutorials, or printed handouts. In one case a group of G and T year 8 students developed and ran a CPD activity for teachers at the ASE annual conference 2010 in Nottingham that also included a question and answer session. Their performance showed a clear improvement in self confidence and team work.

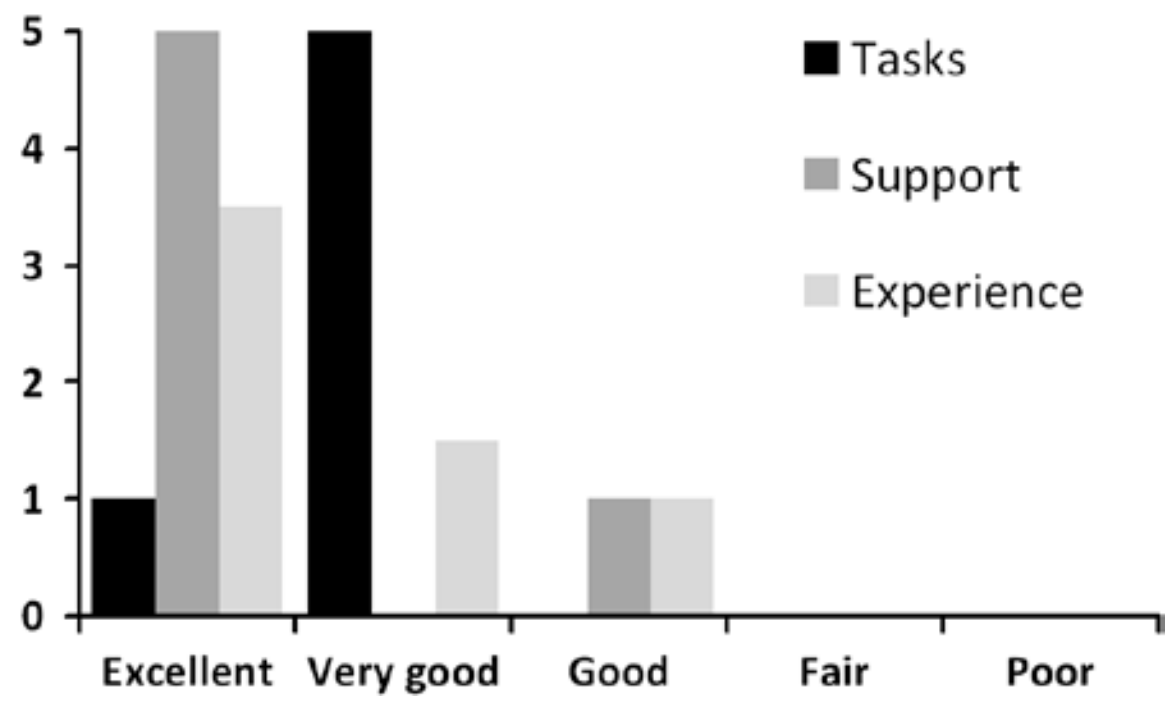

**Figure 7** - The histogram displays the responses of six post-16 work experience students on a Likert scale regarding their rating of tasks given, support provided and the overall experience. The response regarding experience is an average from workplace and university experience. These results are preliminary but indicate a high appreciation of the placement.

## 5. Conclusions

Overall, our Archaeo-Astronomy project has already created a considerable amount of resources and activities that demonstrate how versatile ancient monuments in the Peak District National Park are as outdoor classrooms. Especially, the many links to the school curriculum have been highlighted helping to create a growing motivation and demand for such activities from the schools. This interest has lead to the successful establishment of CPD activities targeting the outdoor classroom for local and international teachers. The created demand is, as outlined in the introduction, the first step in breaking the vicious cycle of neglecting such sites and the outdoor classroom by schools.
Furthermore, many resources have already been developed into considerable useful tools to be piloted in the support of FE/HE learning and teaching activities related to field trips offered at the end of the module. Students will be using Stellarium to encourage their exploration of the historic landscape. Additional impact of our project includes the support of active research work on e.g. rock erosion and light pollution projects involving trainee teachers at NTU.
The most striking result is the success we have had in bridging the gap between universities and schools. Our mini-projects have led to an active collaboration of university staff and school pupils, providing vital feedback for the project and developing a creative and fulfilling work atmosphere for all participants. During the last year the intake of work experience students has doubled. Many of the work experience students follow the development of their projects to see how they have improved and have felt their work contributed to a greater goal.
Ancient monuments e.g. megalithic stone circles offer an ideal learning and teaching environment. Here students can work on project based learning activities and experience a more realistic work environment that awaits them after they leave school.

**Acknowledgements**
I would like to thank all our work experience and placement students that have participated in our project: R. Head, A. Yeung, R. Burton, W. Busby, J. Scott, L. Reszke, N. Sikand-Youngs, R. Lloyd Mills, and J. Allen. Part of this work has included the CPD activities, carried out with the help of T. Sherwood, J. Cantrell, N. Evans, C. Johnson, L. Simpson, and all the participants of the archaeo-astronomy summer school 2009 from The Meden School and Technology College. Further technical support was provided by D. Parker (NTU SST).